\documentclass[prd,preprint,superscriptaddress,amsmath,amssymb,nofootinbib]{revtex4}
\usepackage{graphicx}
\usepackage{dcolumn}
\usepackage{bm}
\usepackage{amssymb}
\usepackage{amsmath}
\usepackage{epsfig}    
\usepackage{color}
\usepackage{slashed}
\usepackage{hhline}

\def\be{\begin{equation}}
\def\ee{\end{equation}}
\newcommand{\bea}{\begin{eqnarray}}
\newcommand{\eea}{\end{eqnarray}}

\begin{document}

\preprint{APCTP Pre2022 - 028}
\preprint{UME-PP-025}
\preprint{KYUSHU-HET-254}

\title{
Neutrinophilic DM annihilation in a model \\
with $U(1)_{L_\mu - L_{\tau}} \times U(1)_{H}$ gauge symmetry
}

\author{Keiko I. Nagao}
\email{nagao@ous.ac.jp}
\affiliation{Department of Physics, Faculty of Science, Okayama University of Science, 1-1 Ridaicho, Okayama, 700-0005, Japan}

\author{Takaaki Nomura}
\email{nomura@scu.edu.cn}
\affiliation{College of Physics, Sichuan University, Chengdu 610065, China}

\author{Hiroshi Okada}
\email{hiroshi.okada@apctp.org}
\affiliation{Asia Pacific Center for Theoretical Physics (APCTP) - Headquarters San 31, Hyoja-dong,
Nam-gu, Pohang 790-784, Korea}
\affiliation{Department of Physics, Pohang University of Science and Technology, Pohang 37673, Republic of Korea}

\author{Takashi Shimomura}
\email{shimomura@cc.miyazaki-u.ac.jp}
\affiliation{Faculty of Education, University of Miyazaki, 
1-1 Gakuen-Kibanadai-Nishi, Miyazaki 889-2192, Japan}
\affiliation{Department of Physics, Kyushu University, 744 Motooka, Nishi-ku, Fukuoka, 819-0395, Japan}

\date{\today}

\begin{abstract}
 We propose a model with two different extra $U(1)$ gauge symmetries; muon minus tauon symmetry $U(1)_{{L_\mu}-L_{\tau}}$ and hidden symmetry $U(1)_H$. Then, we explain muon anomalous magnetic moment, semi-leptonic decays $b\to s\ell\bar\ell$, and dark matter. In particular, we find an intriguing dark matter candidate to be verified by Hyper-Kamiokande and JUNO in the future that request neutrinophilic DM with rather light dark matter mass$\sim{\cal O}(10)$ MeV.
\end{abstract}
\maketitle

\section{Introduction}

An extension of the standard model (SM) of particle physics is required to explain several issues such as the existence of dark matter (DM)  and 
the deviation from the SM prediction of anomalous magnetic dipole of muon (muon $g-2$).
One of the interesting possibilities of extending the SM is an introduction of new $U(1)$ gauge symmetries. 
Such $U(1)$ symmetries can guarantee the stability of DM and interactions mediated by new neutral gauge bosons would explain its relic density.
Furthermore, if new $U(1)$ charges are flavor dependent it is possible to address flavor issues such as muon $g-2$ via new gauge bosons.
From a theoretical viewpoint, multiple extra $U(1)$ symmetries can be induced from theory in high energy scale like the string theories.

In 2021, new result of the muon $g-2$ measurement is reported by the E989 collaboration at Fermilab \cite{Muong-2:2021ojo}: 
\begin{align}
a^{\rm FNAL}_\mu =116592040(54) \times 10^{-11}.
\label{exp_dmu}
\end{align}
Combining the result with the previous BNL one, the muon $g-2$ deviates from the SM prediction by 4.2$\sigma$ level~\cite{Muong-2:2021ojo, Aoyama:2012wk,Aoyama:2019ryr,Czarnecki:2002nt,Gnendiger:2013pva,Davier:2017zfy,Keshavarzi:2018mgv,Colangelo:2018mtw,Hoferichter:2019mqg,Davier:2019can,Keshavarzi:2019abf,Kurz:2014wya,Melnikov:2003xd,Masjuan:2017tvw,Colangelo:2017fiz,Hoferichter:2018kwz,Gerardin:2019vio,Bijnens:2019ghy,Colangelo:2019uex,Blum:2019ugy,Colangelo:2014qya,Hagiwara:2011af} as
\begin{align}
\Delta a^{\rm new}_\mu = (25.1\pm 5.9)\times 10^{-10}.
\end{align}
Although results on the hadronic vacuum polarization (HVP) that are estimated by recent lattice calculations~\cite{Borsanyi:2020mff,Alexandrou:2022amy,Ce:2022kxy} might weaken the necessity of a new physics effect, it is also shown in refs.~\cite{Crivellin:2020zul,deRafael:2020uif,Keshavarzi:2020bfy}~\footnote{The relation between HVP for muon $g-2$ and electroweak precision test is also discussed previously in ref.~\cite{Passera:2008jk}.} that the lattice results imply new tensions with the HVP extracted from $e^+ e^-$ data and the global fits to the electroweak precision observables. 
One of the promising ways to explain the anomaly is introduction of $U(1)_{L_\mu -L_\tau}$ gauge symmetry 
that is often applied for a neutrino mass matrix texture in order to have some predictions as well as
explain the neutrino oscillation data in the lepton sector.    
Here, the associated $Z'$ boson interaction can provide a sizable contribution to muon $g-2$.
In explaining muon $g-2$ anomaly, we need to require the $Z'$ boson from $U(1)_{L_\mu -L_\tau}$ is as light as $\mathcal{O}(10)$ to $\mathcal{O}(100)$ MeV and small gauge coupling of $\mathcal{O}(10^{-4})$ to $\mathcal{O}(10^{-3})$.
On the other hand, if $U(1)_{L_\mu -L_\tau}$ like the $Z'$ boson is heavier than $\mathcal{O}(10)$ GeV, it affects different phenomenology like the lepton flavor non-universality in semi-leptonic $B$ meson decay associated with the process $b \to s \ell^+ \ell^-$~\cite{Altmannshofer:2014cfa, Crivellin:2015mga, Crivellin:2015lwa, Ko:2017yrd, Kumar:2020web,Han:2019diw,Chao:2021qxq,Baek:2019qte,Chen:2017usq,Borah:2021khc,Tuckler:2022fkz,Altmannshofer:2016jzy,Crivellin:2022obd} 
where some deviation from the SM is observed~\cite{Hiller:2003js, Bobeth:2007dw, Aaij:2014ora,Aaij:2019wad, DescotesGenon:2012zf, Aaij:2015oid, Aaij:2013qta,Abdesselam:2016llu, Wehle:2016yoi,Aaij:2017vbb,Aaij:2021vac}.
Although recent LHCb observation regarding this lepton universality is compatible with the SM prediction~\cite{LHCb:2022qnv,LHCb:2022zom}, it is still interesting to investigate the effect of the $Z'$ boson in $b \to s \ell^+ \ell^-$ decays.
Thus we are interested in considering multiple $U(1)_{L_\mu -L_\tau}$ like the $Z'$ bosons whose mass scales are different that provide us richer phenomenological possibilities, 
which can be realized by introducing multiple $U(1)$ gauge symmetries.
Moreover, considering multiple local $U(1)$s, we can have more freedom to accommodate the relic density of DM.

In this paper, we consider a model that has two local $U(1)$ symmetries. One symmetry is $U(1)_{L_\mu -L_\tau}$ and the other one is hidden $U(1)_H$ symmetry where the SM fields are not charged under it.
After the spontaneous breaking of these symmetries, two $Z'$ bosons mix each other, and they can interact with the $\mu$ and $\tau$ flavor leptons in mass basis. 
Then we have two $Z'$ bosons whose masses can be hierarchical in some parameter region inducing different phenomenology.
Both heavy and light $Z'$ can contribute to muon $g-2$ and it is possible to explain the experimental measurements.
The heavy one can affect lepton non-universality in $B$-meson decay when we assume some effective interactions with quarks.
Also, the light one can play a role in explaining DM relic density when we choose light DM as $\mathcal{O}(10)$ MeV.
Interestingly DM annihilation mode is neutrinophilic in our benchmark points and it is safe from current direct and indirect searches constraints.
We can test neutrino signals from DM annihilation in future neutrino experiments where we consider Hyper-Kamiokande(HK) and JUNO as promising candidates.

 This article is organized as follows.
In Sec. II, we introduce our model and show relevant interactions and formulas for phenomenology.
In Sec. III, we show our phenomenological analysis of muon $g-2$, DM physics and neutrino signatures.
 Finally, we devote the summary of our results and the conclusion.

\section{Model setup and phenomenological formulas}

\begin{table}[t!]
\begin{tabular}{|c||c|c|c|c||c|c|c|}\hline\hline  
&  ~$L_L^a$~& ~$e_R^a$~ & ~$\nu_R^a$~ &  ~$\chi$~  &   ~$H$~ & ~$\varphi_1$~ & ~$\varphi_2$~ \\\hline\hline 
$SU(2)_L$ &  $\bm{2}$  & $\bm{1}$ & $\bm{1}$   & $\bm{1}$    & $\bm{2}$  & $\bm{1}$ & $\bm{1}$   \\\hline 
$U(1)_Y$   &  $-\frac12$  & $-1$ & $0$ & $0$ & $\frac12$  & $0$  & $0$ \\\hline
$U(1)_{L_\mu - L_\tau}$  & $\{0,1,-1\}$ & $\{0,1,-1\}$ & $\{0,1,-1\}$ & $0$  & $0$  & $1$  & $1$\\\hline
$U(1)_H$ & $0$ & $0$ & $0$ & $Q_\chi$ & $0$ & $0$ & $1$ \\ \hline
\end{tabular}
\caption{ 
Charge assignments of the fields
under $SU(2)_L\times U(1)_Y\times U(1)_{L_\mu - L_\tau} \times U(1)_H$, where its upper index $a$ is the number of family that runs over $1-3$.}
\label{tab:1}
\end{table}

We propose a model with $U(1)_{L_\mu - L_\tau} \times U(1)_H$ gauge symmetry. 
The SM leptons and right-handed neutrinos $\nu_R$ are charged under $U(1)_{L_\mu - L_\tau}$ where second and third generation ones have charge $1$ and $-1$, respectively.
We also introduce SM singlet Dirac fermion $\chi$ with $U(1)_H$ charge $Q_\chi$, which is our DM candidate.
In the scalar sector, we introduce the SM Higgs field $H$, and the SM singlet scalars $\varphi_1$ and $\varphi_2$.
$H$ is neutral under new gauge symmetry while $\varphi_1$ and $\varphi_2$ have charges $\{1,0\}$ and $\{1,1\}$ under $\{ U(1)_{L_\mu - L_\tau}, U(1)_H\}$ respectively.
All the field contents and their assignments are summarized in Table~\ref{tab:1}.
New terms of Lagrangian and scalar potential are written by
\begin{align}
& \mathcal{L}_{\rm new} = g_{\mu \tau} X_1^\rho (\bar \mu \gamma_\rho \mu - \bar \tau \gamma_\rho \tau + \bar \nu_\mu \gamma_\rho \nu_\mu - \bar \nu_\tau \gamma_\rho \nu_\tau ) 
+ \bar \chi [ i \gamma_\mu (\partial^\mu - i Q_\chi g_H X_2^\mu) - m_\chi ] \chi \nonumber \\
& \qquad \quad + |(\partial_\mu - i g_{\mu \tau} X_{1 \mu}) \varphi_1|^2 +  |(\partial_\mu - i g_{\mu \tau} X_{1 \mu}- i g_{H} X_{2 \mu} ) \varphi_2|^2, \\
& V = \mu_H^2 (H^\dagger H) + \mu_{1}^2 \varphi_1^* \varphi_1 + \mu_{2}^2 \varphi_2^* \varphi_2 + \lambda_H (H^\dagger H)^2 + \lambda_1 (\varphi^*_1 \varphi_1)^2+ \lambda_2 (\varphi^*_2 \varphi_2)^2 \nonumber \\
& \qquad + \lambda_{H \varphi_1} (H^\dagger H) (\varphi^*_1 \varphi_1)  + \lambda_{H \varphi_2} (H^\dagger H) (\varphi^*_2 \varphi_2) +  \lambda_{\varphi_1 \varphi_2} (\varphi^*_1 \varphi_1)  (\varphi^*_2 \varphi_2), 
\end{align}
where $X_1^\mu$ and $X_2^\mu$ are gauge fields of $U(1)_{L_\mu - L_\tau}$ and $U(1)_H$ with corresponding gauge couplings $g_{\mu \tau}$ and $g_H$.

\subsection{Scalar sector \label{sec:scalar}}
In the proposed scenario, we require all the scalar fields develop their vacuum expectation values (VEVs) to break electroweak and $U(1)_{L_\mu - L_\tau} \times U(1)_H$ gauge symmetries denoted by $\langle H\rangle\equiv[0,v/\sqrt2]^T$ and $\varphi_{1,2}\equiv v_{1,2}/\sqrt2$.
Then, the scalar fields are written by
\begin{equation}
H =\left[\begin{array}{c}
w^+\\
\frac{v + \tilde{h} +i z}{\sqrt2}
\end{array}\right],\quad 
\varphi_{1,2}=
\frac{v_{1,2}+ \phi_{1,2} + i z'_{1,2} }{\sqrt2},
\end{equation}
where $w^+$ and $z$ are massless Nambu-Goldstone(NG) bosons which are absorbed by the SM gauge bosons $W^+$ and $Z$, and
linear combinations of $z'_{1,2}$ become NG bosons absorbed by two extra neutral gauge bosons from $U(1)_{L_\mu - L_\tau} \times U(1)_H$.  
The VEVs are obtained by conditions $\frac{\partial {\cal V}}{\partial v} = \frac{\partial {\cal V}}{\partial v_1} =\frac{\partial {\cal V}}{\partial v_2} =0$ that are written by 
\begin{align}
& v \left( \mu_H^2 + \frac{\lambda_{H \varphi_1}}{2} v_1^2 + \frac{\lambda_{H \varphi_2}}{2} v_2^2 \right) + \lambda_H v^3 =0, \\
& v_1 \left( \mu_1^2 + \frac{\lambda_{H \varphi_1}}{2} v^2 + \frac{\lambda_{\varphi_1 \varphi_2}}{2} v_2^2 \right) + \lambda_1 v_1^3 =0, \\
& v_2 \left( \mu_2^2 + \frac{\lambda_{H \varphi_2}}{2} v^2 + \frac{\lambda_{\varphi_1 \varphi_2}}{2} v_1^2 \right) + \lambda_2 v_2^3 =0. 
\end{align}
After the scalar fields developing VEVs, neutral scalars $\tilde h$, $\phi_1$ and $\phi_2$ mix.
We assume the mixings between $\tilde h$ and $\phi_{1,2}$ are small for simplicity and $\tilde h$ is identified as the SM Higgs boson.
Other scalar bosons are not involved in our phenomenological analysis below and we do not discuss further details in this work~\footnote{Collider signatures from a scalar boson decaying into $Z'$ in the context of $U(1)_{L_\mu - L_\tau}$ are discussed in refs.~\cite{Das:2022mmh, Nomura:2020vnk, Nomura:2018yej}. }.

\subsection{Gauge sector \label{sec:gauge}}

After spontaneous symmetry breaking, we obtain mass terms for the gauge fields such that
\begin{align}
|D_\mu \varphi_1|^2 + |D_\mu \varphi_2|^2 \supset 
\frac{1}{2} g^2_{\mu \tau} (v_1^2 + v_2^2) X^\mu_1 X_{1 \mu}  + \frac12 g^2_H v^2_2 X^\mu_2 X_{2 \mu } + g_{\mu \tau} g_H v_2^2 X^\mu_1 X_{2 \mu}.
\end{align}
We thus obtain mass matrix in the basis of $\{X^\mu_1, X^\mu_2 \}$ as 
\begin{equation}
M_{X_1X_2}^2 = \begin{pmatrix} M^2_1 & M^2_{12} \\ M^2_{12} & M^2_2 \end{pmatrix},
\end{equation}
where $M^2_1 \equiv g^2_{\mu \tau} (v_1^2 + v_2^2)$, $M^2_2 \equiv g^2_H v^2_2$ and $M^2_{12} \equiv g_H g_{\mu \tau} v^2_2$.
We can diagonalize the mass matrix by orthogonal matrix, and mass eigenvalues and eigenstates are written by
\begin{align}
& m^2_{Z'_1} = \frac12 (M^2_{1} + M^2_{2}) + \frac12 \sqrt{(M^2_{1} - M^2_{2})^2 + 4 M^4_{12}} \ , \\
& m^2_{Z'_2} = \frac12 (M^2_{1} + M^2_{2}) - \frac12 \sqrt{(M^2_{1} - M^2_{2})^2 + 4 M^4_{12}} \ , \\
& \begin{pmatrix} X_1^\mu \\ X^\mu_2 \end{pmatrix} = 
\begin{pmatrix} \cos \theta_{X} & \sin \theta_X \\ - \sin \theta_X & \cos \theta_X \end{pmatrix}
\begin{pmatrix} Z'^\mu_1 \\ Z'^\mu_2 \end{pmatrix},
\end{align} 
where the mixing angle $\theta_X$ is obtained from
\begin{equation}
\tan 2 \theta_X = \frac{2 M^2_{12}}{M^2_1 - M^2_2}.
\end{equation}
In our scenario, we consider the case of $m_{Z'_1} \gg m_{Z'_2}$ by choosing gauge couplings and VEVs accordingly.
Then $Z'_1$ interaction is applied to explain $B$ decay anomalies while $Z'_2$ dominantly contributes to induce sizable muon $g-2$.

\subsection{Gauge interaction in mass basis}

In mass basis, we write relevant new gauge interactions as 
\begin{align}
\label{eq:gauge-int}
\mathcal{L} \supset \ & g_{\mu \tau} (c_X Z'^\mu_1 + s_X Z'^\mu_2) (\bar \mu \gamma_\mu \mu - \bar \tau \gamma_\mu \tau) + g_{\mu \tau} (c_X Z'^\mu_1 + s_X Z'^\mu_2) (\overline{ \nu_\mu} \gamma_\mu P_L \nu_\mu - \overline{ \nu_\tau} \gamma_\mu P_L \nu_\tau) \nonumber \\
& + Q_\chi g_H (- s_X Z'^\mu_1 + c_X Z'^\mu_2) \bar \chi \gamma_\mu \chi,
\end{align}
where $c_X(s_X) = \cos \theta_X (\sin \theta_X)$.

Our $Z'$ bosons can decay into leptons $\{ \mu \bar \mu, \tau \bar \tau, \nu_\mu \bar \nu_\mu, \nu_\tau \bar \nu_\tau \}$ and $\chi \bar \chi$ when these modes are kinematically allowed.
Partial decay widths of the $Z'_1$ and $Z'_2$ bosons are given by 
\begin{align}
\label{eq:ZpWidth}
\Gamma(Z'_{1} \to \ell' \bar \ell') & = \frac{c_X^2 g^2_{\mu \tau}}{8 \pi} m_{Z'_1} \left( 1 + \frac{4}{3} \frac{m_{\ell'}^2}{m^2_{Z'_1}} \sqrt{1 - \frac{4 m^2_{\ell'} }{m^2_{Z'_1}} } \right), \nonumber \\
\Gamma(Z'_{1} \to \nu_{\ell'} \bar \nu_{\ell'}) & = \frac{c_X^2 g^2_{\mu \tau}}{8 \pi} m_{Z'_{1} }, \nonumber \\  
\Gamma(Z'_{1} \to \chi \bar \chi) & = \frac{Q_\chi^2 s_X^2 g^2_{H}}{8 \pi} m_{Z'_1} \left( 1 + \frac{4}{3} \frac{m_{\chi}^2}{m^2_{Z'_1}} \sqrt{1 - \frac{4 m^2_{\chi} }{m^2_{Z'_1}} } \right), \nonumber \\
\Gamma(Z'_{2} \to \ell' \bar \ell') & =  \frac{s_X^2 g^2_{\mu \tau}}{8 \pi} m_{Z'_2} \left( 1 + \frac{4}{3} \frac{m_{\ell'}^2}{m^2_{Z'_2}} \sqrt{1 - \frac{4 m^2_{\ell'} }{m^2_{Z'_2}} } \right), \nonumber \\
\Gamma(Z'_{2} \to \nu_{\ell'} \bar \nu_{\ell'}) & =  \frac{s_X^2 g^2_{\mu \tau}}{8 \pi} m_{Z'_{2} }, \nonumber  \\
\Gamma(Z'_{2} \to \chi \bar \chi) & =  \frac{Q_\chi^2 c_X^2 g^2_{H}}{8 \pi} m_{Z'_2} \left( 1 + \frac{4}{3} \frac{m_{\chi}^2}{m^2_{Z'_2}} \sqrt{1 - \frac{4 m^2_{\chi} }{m^2_{Z'_2}} } \right), 
\end{align} 
where $\ell' = \{\mu, \tau\}$.
When the $Z'_{1,2}$ boson mass is $\mathcal{O}(10)$ GeV to $\mathcal{O}(100)$ GeV, their couplings and masses are strongly constrained by 
LHC data searching for $pp \to \mu^+ \mu^- Z'(\to \mu^+ \mu^-)$ signal~\cite{CMS:2018yxg}. We can also find other constraints in the context of $L_{\mu}-L_{\tau}$ scenario in ref.~\cite{Chun:2018ibr} which are subdominant in the scenario of this work.
The constraint can be relaxed when $Z'_{1,2}$ dominantly decays into $\chi \bar \chi$. 
We will take this constraint into account in our numerical analysis below.

\subsection{Muon $g-2$}

In our model, both the $Z'_1$ and $Z'_2$ bosons contribute to muon $g-2$ since they interact with muon as given by Eq.~\eqref{eq:gauge-int}.
Calculating one-loop diagrams, we obtain muon $g-2$ such that
\begin{equation}
\Delta a_\mu = \frac{g^2_{\mu \tau} m^2_\mu}{4 \pi} \int_0^1 dx  \left[ c_X^2 \frac{x^2 (1-x)}{x^2 + (1-x) r_{Z'_1}} + s_X^2 \frac{x^2 (1-x)}{x^2 + (1-x) r_{Z'_2}}  \right],
\end{equation}
where $r_{Z'_{1(2)}} \equiv m_{Z'_{1(2)}}^2/m^2_\mu $.
We apply the $3 \sigma$ region obtained by combining FNAL and BNL results that require new contribution to satisfy
\begin{equation}
(25.1- 3\times5.9)\times 10^{-10} \leq \Delta a^{\rm new}_\mu \leq (25.1 + 3 \times 5.9)\times 10^{-10}. \label{eq:gm2}
\end{equation}
This constraint is imposed in our numerical calculation below.

\subsection{Constraint from neutrino trident}
Neutrino trident production processes, $\nu + N \to \nu + N + \mu + \bar{\mu}$, are neutrino scattering off nucleus producing a muon and anti-muon pair. The cross section of the muon neutrino trident has been measured by past experiments, CHARM-II \cite{CHARM-II:1990dvf}, CCFR \cite{CCFR:1991lpl} and NuTeV \cite{NuTeV:1999wlw}. The current limits on the total cross section are respectively given by 
\begin{align}
\frac{\sigma_{\mathrm{CHARM-II}}}{\sigma_{\mathrm{SM}}} &= 1.58 \pm 0.57, \\
\frac{\sigma_{\mathrm{CCFR}}}{\sigma_{\mathrm{SM}}} &= 0.82 \pm 0.28, \\ 
\frac{\sigma_{\mathrm{NuTeV}}}{\sigma_{\mathrm{SM}}} &= 0.72^{+1.73}_{-0.72}, 
\end{align}
where $\sigma_{\mathrm{SM}}$ is the SM prediction \cite{Belusevic:1987cw}\footnote{The cross sections were calculated in the V-A theory in Refs.\cite{Czyz:1964zz, Lovseth:1971vv,Fujikawa:1973vu,Koike:1971tu,Koike:1971vg,Brown:1972vne}.}. 

It was shown in  \cite{Altmannshofer:2014pba} that these limits severely constrain the $Z'$ mass and its coupling to neutrinos. The sensitivity in on-going and future experiments have been studied, e.g. in \cite{Kaneta:2016uyt, Araki:2017wyg, Magill:2016hgc, Ge:2017poy, Falkowski:2018dmy, Altmannshofer:2019zhy,Shimomura:2020tmg}. The trident cross section of this model can 
be obtained by simply replacing the coupling of the SM as
\begin{align}
    g_{L/R} \to g_{L/R} - \frac{\sqrt{2}g_{\mu \tau}^2}{4G_F}  \sum_i \frac{a_i}{q^2 - m^2_{Z_i'}},
\end{align}
where $a_1 = c_X^2,~a_2=s_X^2$ and $g_{L/R}$ is 
\begin{align}
    g_L &= \frac{1}{2} + \sin^2\theta_W, \\
    g_R &=  \sin^2\theta_W.
\end{align}
The concrete forms of the amplitudes are given in \cite{Shimomura:2020tmg}. 

To obtain the constraints from the tridents, we have calculated the cross section for CHARM-II and CCFR and compared them with the SM predictions\footnote{The result from NuTeV has large uncertainty and includes the null result. Therefore we do not use this result.}.

\subsection{Effective interactions for $B \to K^{(*)} \ell^+ \ell^-$ and $B_s$--$\overline{B_s}$ mixing}

\begin{figure}[tb]
\begin{center}
\includegraphics[width=65.0mm]{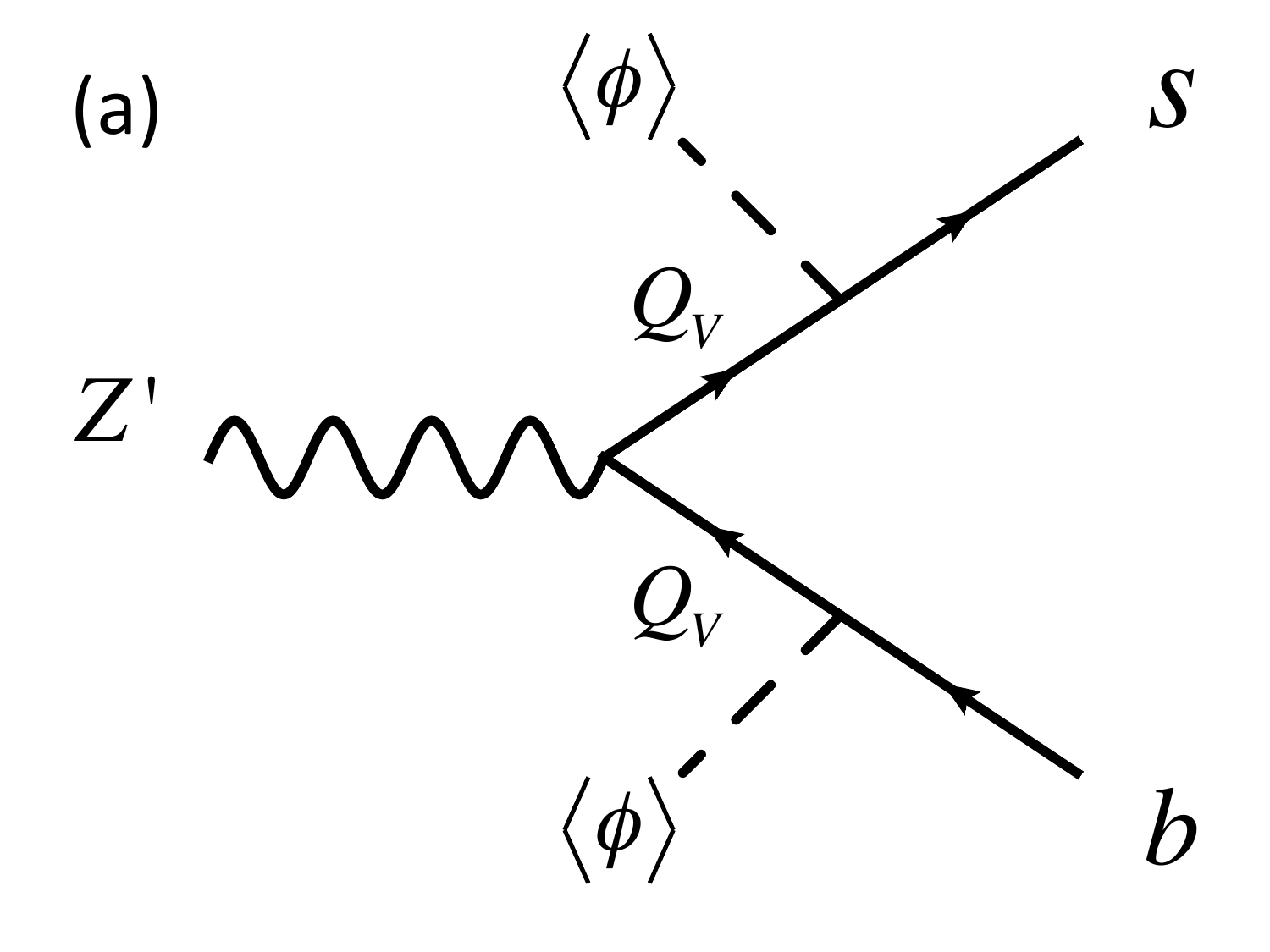} \quad
\includegraphics[width=65.0mm]{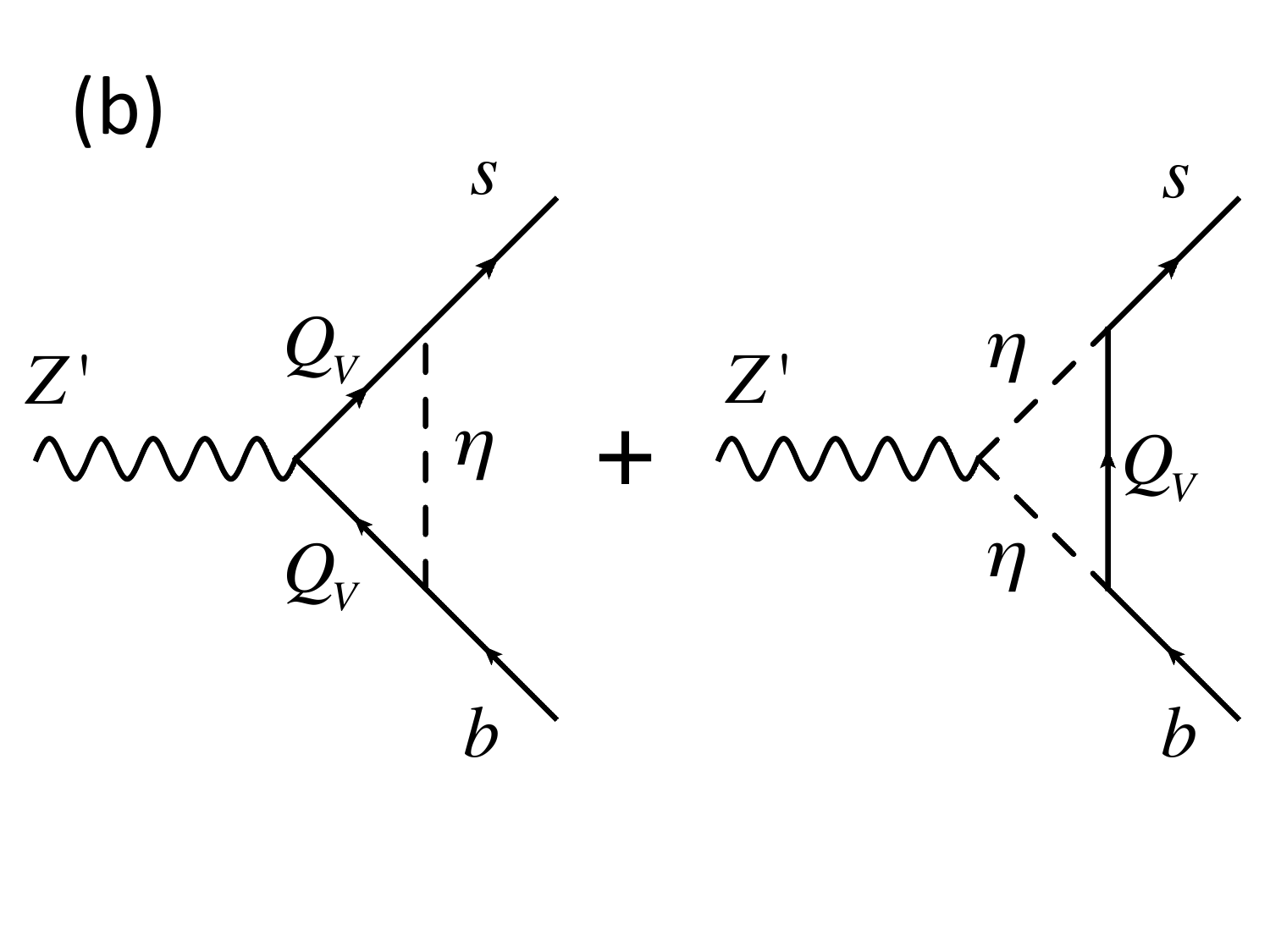}
\caption{(a) A diagram inducing $Z'_\mu \bar b \gamma^\mu P_L s$ interaction via mixing between the SM b(s) quark and vector-like quark(s) $Q_V$ with new $U(1)$ charge where scalar $\phi$ connect them. 
(b) A diagram inducing the interaction via one-loop effect where $\eta$ is a scalar field with new $U(1)$ charge. }
  \label{fig:diagram}
\end{center}\end{figure}

Firstly we review the mechanisms to explain B decay anomalies by $Z'$ boson from local $U(1)_{L_\mu - L_\tau}$. 
The anomalies can be explained if we have a quark flavor violating interaction $Z'_\mu \bar b \gamma^\mu P_L s$ and $Z'$-muon interaction $Z'_\nu \bar \mu \gamma^\nu \mu$.
These interactions induce an effective interaction 
\begin{equation}
\mathcal{H}_{\rm eff} = - \frac{4 G_F}{\sqrt{2}} V_{tb} V^*_{ts} \frac{\alpha}{4 \pi} C_9 (\bar b \gamma^\nu P_L s)(\bar \mu \gamma_\nu \mu),
\end{equation}
where the $C_9$ is a corresponding Wilson coefficient, $G_F$ is the Fermi constant, $V_{tb}$ and $V_{ts}$ are elements of CKM matrix, and $\alpha$ is the fine structure constant. 
To induce $Z'_\mu \bar b \gamma^\mu P_L s$, we need new fields since the $Z'$ boson from $U(1)_{L_\mu - L_\tau}$  does not interact with quarks.
Possible ways to induce such a flavor violating interaction can be achieved as follows.
\begin{enumerate}
\item $Z'_\mu \bar b \gamma^\mu P_L s$ interaction is induced via mixing between $U(1)_{L_\mu - L_\tau}$  charged vector-like quark $Q_V$ and the SM ones~\cite{Altmannshofer:2014cfa}. 
We also need a new scalar field $\phi$ with $U(1)_{L_\mu - L_\tau}$  charge to connect a vector-like quark and SM ones after developing its VEV. An example of diagram inducing the interaction by mixing effect 
is shown in Fig.~\ref{fig:diagram}(a).
\item The interaction is induced at one-loop level by introducing a vector-like quark $Q_V$ and new scalar field $\eta$ that are charged under $U(1)_{L_\mu - L_\tau}$~\cite{Ko:2017yrd,Chen:2017usq}. 
An example of diagram inducing the interaction by radiative correction is shown in Fig.~\ref{fig:diagram}(b). Interestingly, a scalar field $\eta$ can be DM candidate.
\end{enumerate}
In fact, they can be obtained by the same field contents, a vector-like quark $Q_V$ and a scalar field ($\phi = \eta$) charged under a new $U(1)$ gauge group(s), 
where the difference is whether new scalar develops a VEV or not.
In this paper, we do not discuss details and we just assume effective  $X_{1(2) \mu} \bar b \gamma^\mu P_L s$ couplings induced by one of such mechanisms.

In our analysis, we just write effectively induced interactions by
\begin{equation}
\label{eq:effective_bs}
\mathcal{L}_{\rm eff} = (g^L_{bs X_1} X^\mu_1 +g^L_{bs X_2} X^\mu_2 ) \overline{b_L} \gamma_\mu s_L,
\end{equation}
where we consider couplings $g^L_{bs X_{1,2}}$ are free parameters.
In mass basis of the gauge bosons $\{Z'^\mu_1, Z'^\mu_2 \}$, the interaction is rewritten by
\begin{align}
\label{eq:bsZp}
\mathcal{L}_{\rm eff} &= \left[ (g^L_{bs X_1} c_X - g^L_{bsX_2} s_X) Z'^\mu_1 +  (g^L_{bs X_1} s_X + g^L_{bsX_2} c_X) Z'^\mu_2 \right] \overline{b_L} \gamma_\mu s_L \nonumber \\
& \equiv \left[ g^L_{bs Z'_1}  Z'^\mu_1 +  g^L_{bsZ'_2} Z'^\mu_2 \right] \overline{b_L} \gamma_\mu s_L.
\end{align}
Hereafter we consider $g^L_{bs Z'_1}$ and $g^L_{bs Z'_2}$ as free parameters.

From $Z'$ exchanging diagram, we obtain contribution to the Wilson coefficient $C_9$ that is required to explain $B$ anomalies.
Here we assume $g^L_{bs Z'_1} \gg g^L_{bs Z'_2}$ that is required by constraints from $B \to K^{(*)} Z'$ decay as can be seen in the next subsection.
Then contribution to $C_9$ from $Z'_1$ exchange is given by
\begin{equation}
\label{eq:C9Zp}
C_9^{Z'} \simeq - \frac{ \pi c_X g_{\mu \tau} g^L_{bs Z'_1}}{\sqrt{2} V_{tb} V^*_{ts} \alpha G_F m_{Z'_1}^2}. 
\end{equation}

{\it Constraint from $B_s$--$\overline{B_s}$ mixing} :  the effective interactions in Eq.~\eqref{eq:effective_bs} also induce the mixing between $B_s$ and $\overline{B_s}$ mesons, 
and a constraint from the meson mixing should be considered.
We can find the ratio of $B_s$--$\overline{B_s}$ mixing between SM and $Z'$ contributions as~\cite{Tuckler:2022fkz,Altmannshofer:2016jzy} 
\begin{equation}
\frac{M^{Z'}_{12}}{M^{\rm SM}_{12}} \simeq \frac{(g^L_{bs Z'_1})^2}{m_{Z'_1}^2} \frac{16 \sqrt2  \pi^2}{g^2 G_F (V_{tb} V^*_{ts} )^2 S_0},
\end{equation} 
where $S_0 \simeq 2.3$ is the SM loop function~\cite{Inami:1980fz,Buchalla:1995vs}.
Note that we ignored a contribution from $Z'_2$ since we adopt assumption of $g^L_{bs Z'_1} \gg g^L_{bs Z'_2}$.
Then we rewrite $g^L_{bs Z'_1}$ in terms of $C_9^{Z'}$ from Eq.~\eqref{eq:C9Zp} as follows:
\begin{equation}
g^L_{bs Z'_1} = \sqrt{2} V_{tb} V^*_{ts} \frac{\alpha}{\pi} G_F C_9^{Z'} \frac{m_{Z'_1}^2}{c_X g_{\mu \tau}}.
\end{equation}
Requiring $|M^{Z'}_{12}|/|M^{{\rm SM}}_{12}| < 0.12$~\cite{Charles:2020dfl}, we obtain the constraint
\begin{equation}
m_{Z'_1} < 2.1 \ {\rm TeV} \times \frac{c_X g_{\mu \tau}}{|C_9^{Z'}|}.
\end{equation}
We apply the constraint assuming $C_9^{Z'}=-1.01$ that is the best-fit value in our numerical analysis below.

\subsection{$B \to K^{(*)} Z'$ decay width}

In our scenario we consider light $Z'_2$ to explain muon $g-2$, and we should take into account constraints from $B \to K^{(*)}Z'$ decay processes	 
that are induced by quark flavor violating coupling in Eq.~\eqref{eq:bsZp}.
We can estimate decay widths such that~\cite{Crivellin:2022obd}
\begin{align}
& \Gamma(B \to K Z'_2) = \frac{(g^L_{sb Z'_2})^2 f^2_+(m_{Z'_2}^2)}{16 \pi m_B^2 m^2_{Z'_2}} \lambda(m_B^2, m_K^2, m_{Z'_2}^2)^{\frac32},  \\
& \Gamma(B \to K^* Z'_2) \nonumber \\
&= \frac{(g^L_{sb Z'_2})^2 V^2(m^2_{Z'})}{8 \pi m_B^2 (m_B + m_{K^*})^2 } \lambda(m_B^2, m_{K^*}^2, m_{Z'}^2)^{\frac32} 
+  \frac{(g^L_{sb Z'_2})^2 A_2^2(m^2_{Z'_2})}{64 \pi m_B^2 m_{K^*}^2 m^2_{Z'} (m_B + m_{K^*})^2 } \lambda(m_B^2, m_{K^*}^2, m_{Z'_2}^2)^{\frac52} \nonumber \\
&+  \frac{(g^L_{sb Z'_2})^2 A_1^2(m^2_{Z'_2}) [m_B^4+m^4_{K^*} +m^4_{Z'}+10 m^2_{Z'_2} m^2_{K^*} -2 m^2_B (m^2_{K^*} + m^2_{Z'_2})] }{64 \pi m_B^2 m_{K^*}^2 m_{Z'_2}^2} \lambda(m_B^2, m_{K^*}^2, m_{Z'_2}^2)^{\frac12} \nonumber \\
&+ \frac{(g^L_{sb Z'_2})^2 A_1(m^2_{Z'_2}) A_2(m^2_{Z'_2})(m_B^2 - m^2_{K^*}-m^2_{Z'_2})}{32 \pi m_B^2 m_{K^*}^2 m^2_{Z'_2}} \lambda(m_B^2, m_{K^*}^2, m_{Z'_2}^2)^{\frac32},
\end{align}
where $\lambda(x,y,z) = x^2 +y^2+z^2 - 2xy-2xz-2yz$ and $\{f_+, V, A_1, A_2 \}$ are form factors for $B$ and $K^{(*)}$ mesons~\cite{Bailey:2015dka}.
In our scenario, we require $m_{Z'_2} < 2 m_\mu$ and $Z'_2$ decays into neutrinos or DM that are not observed by the detectors at the experiments searching for rare $B$ decay modes.
We note that narrow width approximation is relevant in our case for $Z'_2$ decay. We consider $Z'_2$ decaying into only neutrinos in considering the constraint and decay width is narrow
since corresponding coupling is $g_{\mu \tau} \sin \theta_X \lesssim 0.5 \times 0.1$ in our scenario. 
We then adopt the following experimental constraints~\cite{Belle:2017oht, Belle-II:2021rof}
\begin{align}
& BR(B \to K \nu \bar \nu) < 1.6 \times 10^{-5}, \qquad BR(B \to K^* \nu \bar \nu ) < 2.7 \times 10^{-5}, \nonumber \\
& BR(B^+ \to K^+ \nu \bar \nu) < 4.1 \times 10^{-5}.
\end{align}
We find that constraint becomes stronger when we take $q^2$ dependence of detector efficiency into account for $B^+ \to K^+ \nu \bar \nu$ search~\cite{Belle-II:2021rof} 
since the efficiency is higher for small $q^2$ region ($q^2 = m_{Z'_2}^2 \lesssim (0.2 \ {\rm GeV})^2$).
But we find that the strongest upper bounds of effective coupling $g^L_{bsZ'_2}$ for our benchmark points are obtained from constraint on $BR(B \to K^* \nu \bar \nu )$ 
when we assume efficiency for $B \to K^{*} Z'(\to \nu \bar \nu)$ is the same as that of $B \to K^{*} \bar \nu \nu$ in the SM since $q^2$ dependence other than $B^+ \to K^+ \nu \bar \nu$ is not known~\footnote{The constraint could be stronger when we take $q^2$ dependence into account since 
efficiency at small $q^2$ region  could be higher as in the $B^+ \to K^+ \nu \bar \nu$ search.}.

\section{Numerical analysis and phenomenological results}
 
 In this section, we perform numerical analysis to search for allowed parameter sets that can accommodate required values of muon $g-2$ and $C_9^{Z'}$, evading phenomenological constraints. 
 We then discuss DM physics of the model for benchmark points from allowed parameter sets.
 
 \subsection{Parameter scan for muon $g-2$ and $B$ anomalies}
 
\begin{figure}[tb]
\begin{center}
\includegraphics[width=65.0mm]{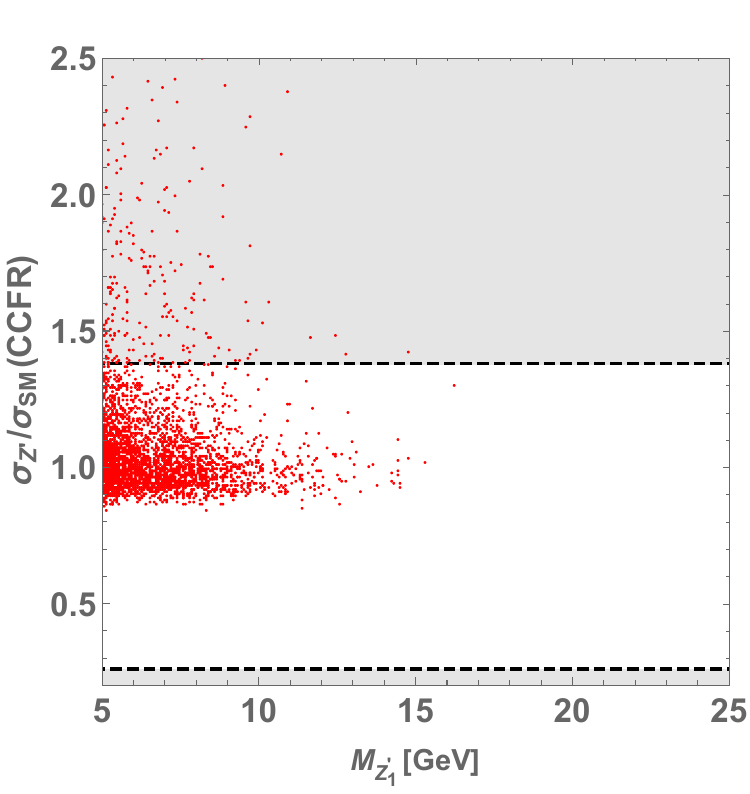} \quad
\includegraphics[width=65.0mm]{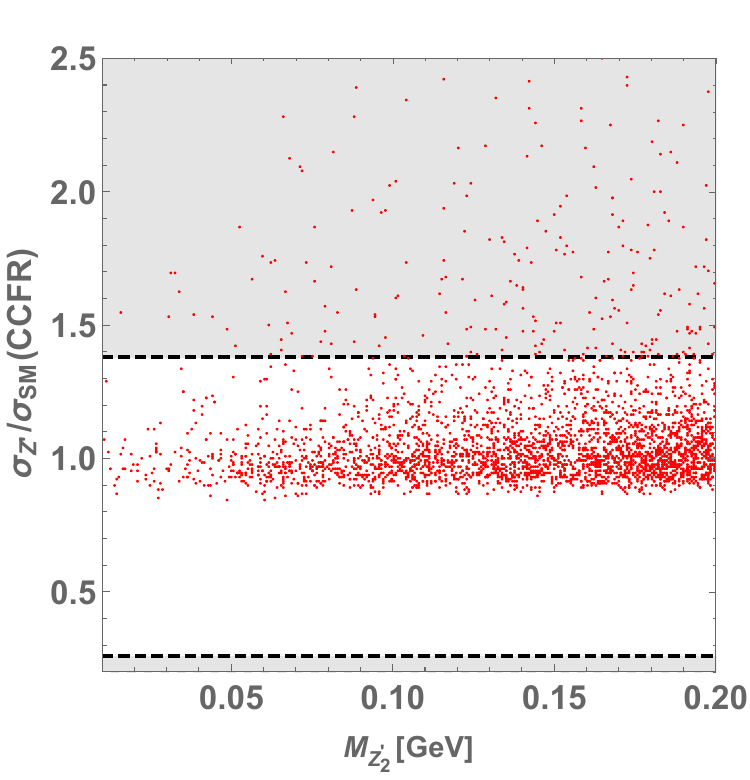}
\includegraphics[width=65.0mm]{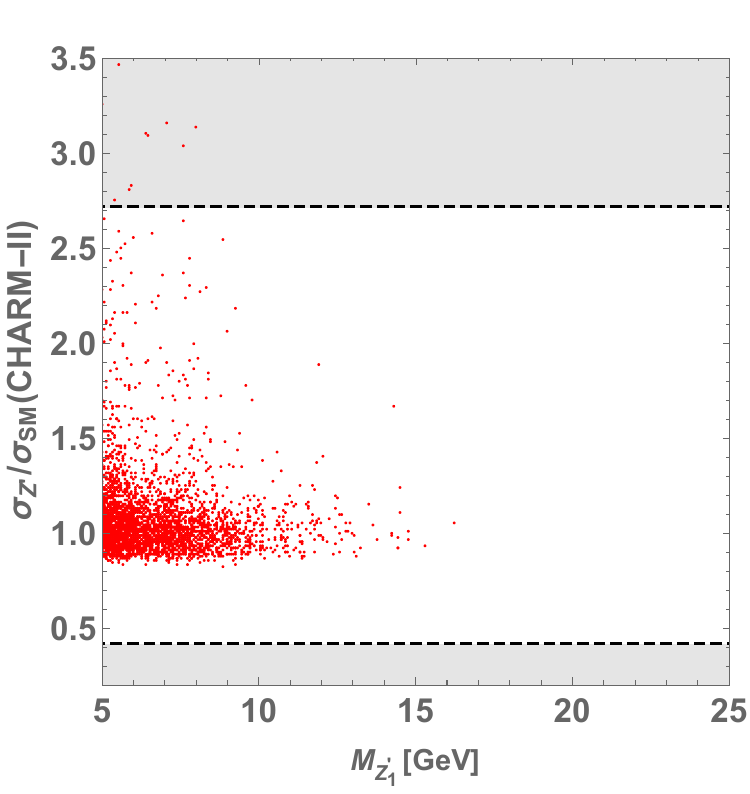} \quad
\includegraphics[width=65.0mm]{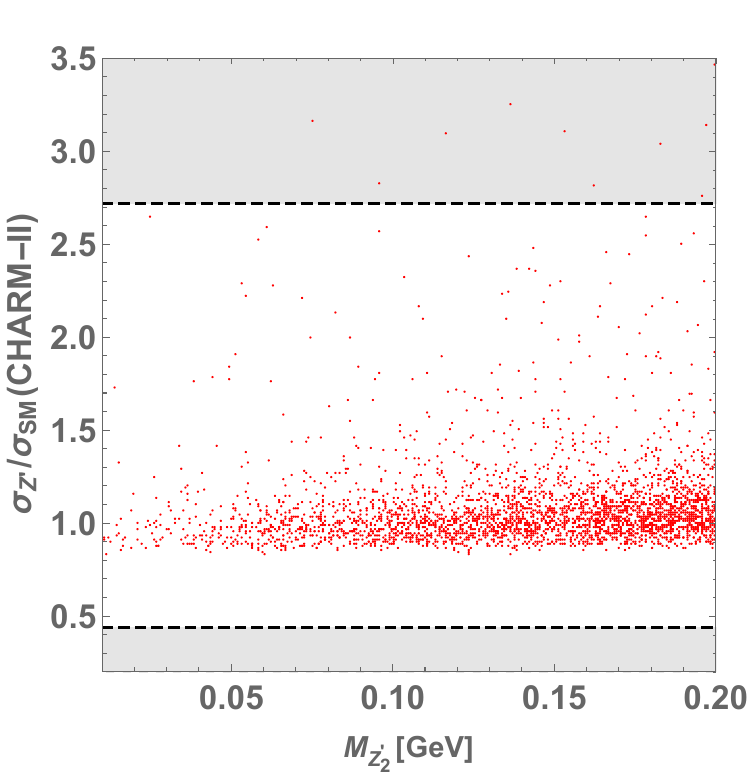}
\caption{The ratio of neutrino trident production cross sections between the SM one and the one including both SM and $Z'$ contributions. The left(right) plots express cross section ratio as functions of $Z'_{1(2)}$ masses. The upper(lower) plots correspond to the ratio in CCFR(CHARM-II) experiments. The shaded regions are excluded by the experiments. }
  \label{fig:trident}
\end{center}\end{figure}

In our analysis, relevant free parameters are 
\begin{equation}
\{ g_{\mu \tau}, g_{H}, v_1, v_2 \},
\end{equation}
and our outputs are $\{ m_{Z'_1}, m_{Z'_2}, \sin \theta_X, \Delta a_\mu \}$ obtained from formulae in the previous section.
The charge $Q_X$ for DM is chosen to be $1$ for simplicity.
We scan our free parameters in the following range:
\begin{align}
g_{\mu \tau} \in [10^{-3}, 10^{-1}], \quad g_{H} \in [10^{-3}, 0.5], \quad v_1 \in [10^{2}, 10^{4}] \ {\rm GeV}, \quad v_2 \in [1, 10^{3}] \ {\rm GeV}.
\end{align}
We then require the following conditions in addition to Eq.~\eqref{eq:gm2};
\begin{align}
& m_{Z'_1} > 5 \ {\rm GeV}, \quad m_{Z'_2} < 0.2 \ {\rm GeV}, \label{eq:zpMass-cond}
\end{align}
where the first constraint is motivated to explain $b \to s \ell^+ \ell^-$ anomalies and the second constraint is imposed to realize DM annihilation into neutrinos. 
In addition, we require $Z'_{1}$ dominantly decaying into $\chi \bar \chi$ by making $\Gamma(Z'_1 \to \chi \bar \chi) \gg \Gamma(Z'_1 \to \ell' \bar \ell', \nu_\ell \bar \nu_{\ell'})$ in Eq.~\eqref{eq:ZpWidth} imposing the condition
\begin{equation}
\label{eq:const2}
\frac{g_{\mu \tau} \cos \theta_X}{g_H \sin \theta_X} < 3, 
\end{equation}
in order to relax the constraint from the LHC search for
$pp \to \mu^+ \mu^- Z'(\to \mu^+ \mu^-)$ signal.

In Fig.~\ref{fig:trident}, we show the ratio of neutrino trident cross section between the SM one and the one including contributions from both the SM and $Z'$ bosons $\sigma_{Z'}$. The plots in the left(right) sides express the ratio as functions of $Z'_{1(2)}$ mass. The upper(lower) plots correspond to the ratio in CCFR(CHARM-II) experiments. The shaded region is excluded by the experiments by 90 $\%$ confidence level (CL). We find some parameter points are excluded where lighter $Z'_1$ region and/or heavier $Z'_2$ region tend to be constrained inducing a larger ratio. 

\begin{figure}[tb]
\begin{center}
\includegraphics[width=65.0mm]{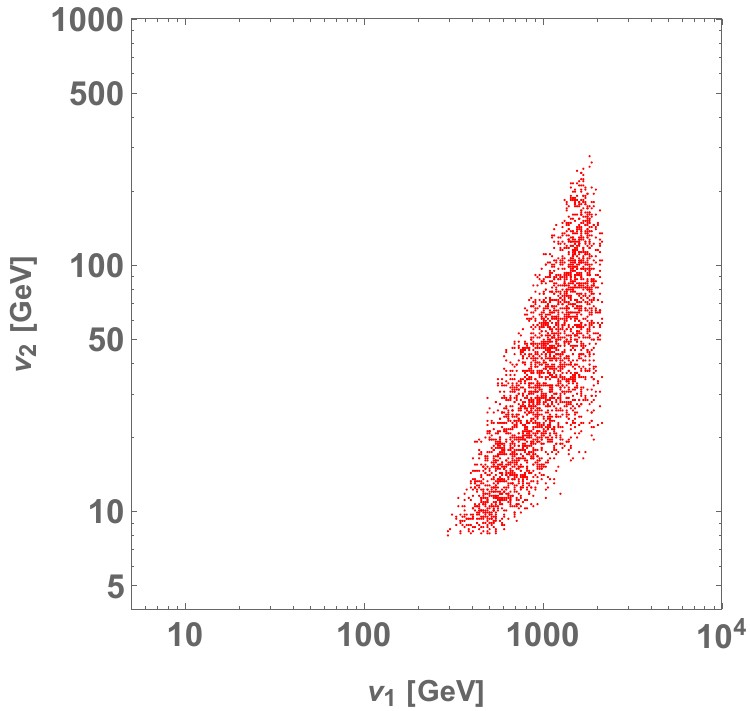} \quad
\includegraphics[width=65.0mm]{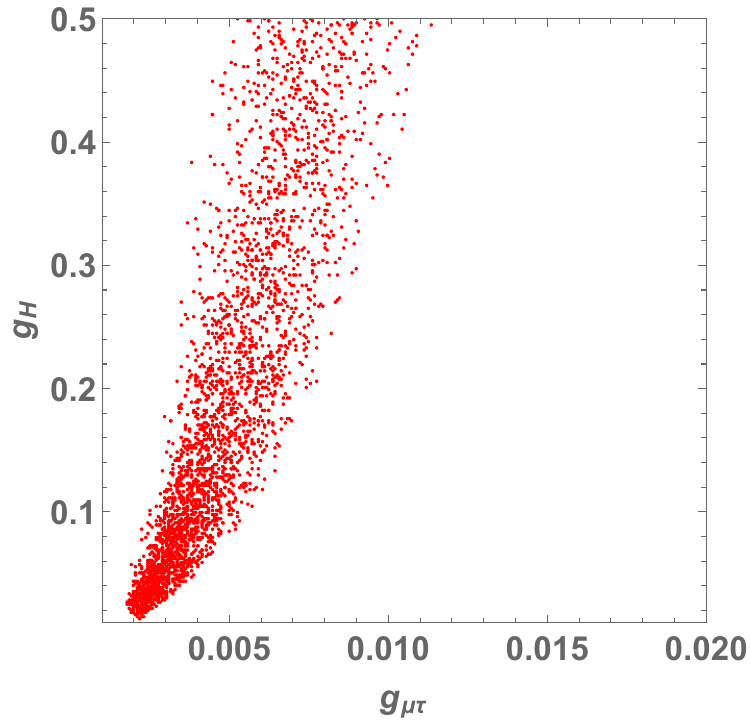}
\caption{Left: correlation between $v_1$ and $v_2$ for our allowed region. Right: correlation between $g_{\mu \tau}$ and $g_H$.}
  \label{fig:correlation}
\end{center}\end{figure}

In the left panel of Fig.~\ref{fig:correlation}, we show correlation between $v_{1}$ and $v_{2}$ for our allowed parameter region.
In addition, the right panel of Fig.~\ref{fig:correlation} shows the correlation between two new gauge couplings.
We find some correlation between them when we require masses of $Z'_{1,2}$ as Eq.~\eqref{eq:zpMass-cond} and muon $g-2$ within $3 \sigma$ CL.

\begin{figure}[tb]
\begin{center}
\includegraphics[width=65.0mm]{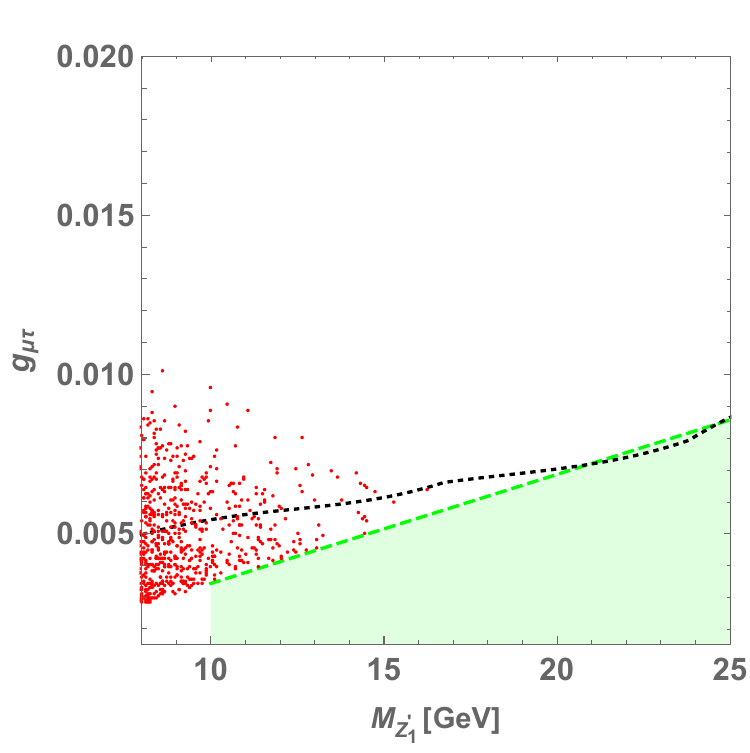} \quad
\includegraphics[width=65.0mm]{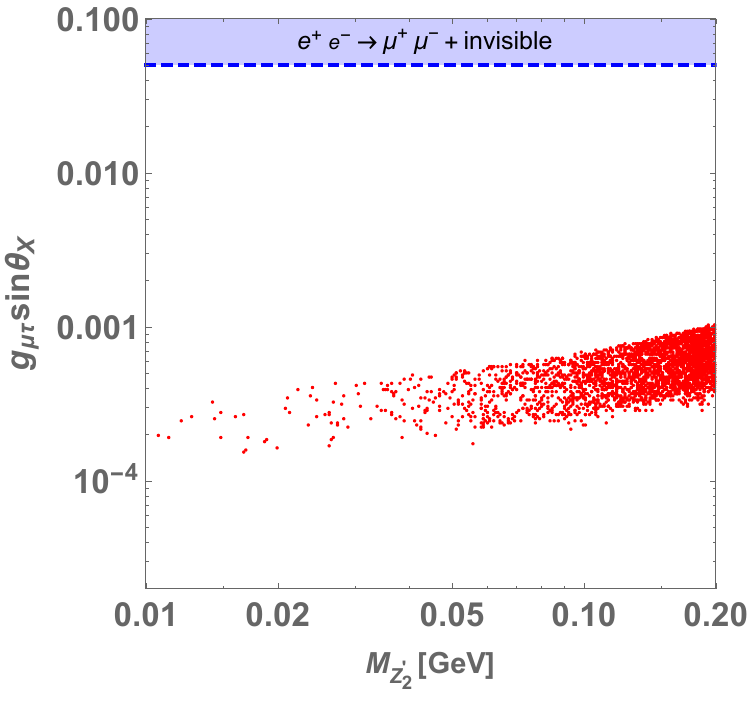}
\caption{The left(right) plot: allowed parameter region that can induce sizable muon $g-2$ within 3$\sigma$ range, on $\{m_{Z'_1}, g_{\mu \tau}\}$ ($\{m_{Z'_2}, g_{\mu \tau} \sin \theta_X \}$) plane. The green-shaded region is disfavored when we explain $b \to s \ell^+ \ell^-$ anomalies. The dotted curve in the left plot is the LHC constraint when $Z'_1$ cannot decay into the DM. The blue-shaded region in right panel is excluded region from ``$e^+ e^- \to \mu^+ \mu^- + \text{invisible}$" search at the Belle II experiment.} 
  \label{fig:parameter}
\end{center}\end{figure}

In left(right) plots of Fig.~\ref{fig:parameter}, we show the allowed parameter region that satisfies the constraints from neutrino trident and induces sizable muon $g-2$ within 3$\sigma$ CL, on $\{m_{Z'_1}, g_{\mu \tau}\}$ ($\{m_{Z'_2}, g_{\mu \tau} \sin \theta_X \}$) plane.
The green shaded region in the left plot is disfavored when we explain $b \to s \ell^+ \ell^-$ anomalies via $Z'_1$ interaction. 
Also, for comparison, we added a dotted curve in the same plot indicating the LHC constraint from 
$pp \to \mu^+ \mu^- Z'(\to \mu^+ \mu^-)$ signal search when $Z'_1$ does not decay into DM; this is not relevant in our case since $Z'_1$ dominantly decays into DM under our requirement in Eq.~\eqref{eq:const2}.
We also show excluded region from ``$e^+ e^- \to \mu^+ \mu^- + \text{invisible}$" search at the Belle II experiment~\cite{Belle-II:2019qfb} as blue shaded region in the right panel of Fig.~\ref{fig:parameter}~\footnote{There is another constraint from ``$pp \to \mu^+ \mu^- +\text{anything}$" search~\cite{Bishara:2017pje} when $Z'_{1,2}$ mass is few GeV scale. We omit the excluded region since the mass region is outside the plots in Fig.~\ref{fig:parameter}}. 
Note also that more parameter region can be tested searching for the ``$e^+ e^- \to \mu^+ \mu^- + \text{invisible}$" process at the future LHC experiments~\cite{Elahi:2015vzh}; 
 analysis in the reference indicates that coupling region of $g_{\mu \tau} \sin \theta_X \gtrsim 3 \times 10^{-4}$ can be tested for $m_{Z'_2} < 2 m_\mu$ at the HL-LHC with 3 ab$^{-1}$ integrated luminosity.

In order to discuss DM physics, we choose several benchmark points from our allowed parameter region.
In Table~\ref{tab:2}, we summarize benchmark points (BPs) that satisfy $\Delta a_\mu$ within $1\sigma$ C.L. where we also show effective coupling $g_{bs Z'_1}$ giving $C_9 ^{Z'}= -1.01$ 
and the upper limit of $g_{bs Z'_2}$ avoiding $B$ meson decay constraints.

\begin{table}[t!]
\begin{tabular}{|c|cccc|cccccc|}\hline
 & \multicolumn{4}{|c|}{Input}  & \multicolumn{6}{|c|}{Output} \\ \hline
 & $g_{\mu \tau}$ & $g_H$ & $v_1/{\rm GeV}$ & $v_2/{\rm GeV}$ & $m_{Z'_1}/{\rm GeV}$ & $m_{Z'_2}/{\rm GeV}$ & $\sin \theta_X$ & $\Delta a_\mu$ & $g_{bs Z'_1}^{(C_9 = -1.01)}$ & $g_{bs Z'_2}^{\rm max}$ \\ \hline
 BP1 & 0.0048 & 0.16  & 1.7 $\times 10^3$ & 47 & 11 & 0.19 & 0.15 & 2.4$\times 10^{-9}$  & $3.9 \times 10^{-5}$ & $5.6 \times 10^{-11}$ \\
 BP2 & 0.0062 & 0.30 & 1.3 $\times 10^3$ & 26 & 11 & 0.18 & 0.11 & 2.1 $\times 10^{-9}$  & $3.1 \times 10^{-5}$ & $5.3 \times 10^{-11}$ \\
 BP3 & 0.0072 & 0.33 & 1.2 $\times 10^3$ & 24 & 12 & 0.19 & 0.12 & 3.1 $\times 10^{-9}$   & $3.1 \times 10^{-5}$ & $5.6 \times 10^{-11}$   \\
 BP4 & 0.0056 & 0.35 & 1.8 $\times 10^3$ & 28 & 14 & 0.20 & 0.11 & 1.5 $\times 10^{-9}$ & $5.4 \times 10^{-5}$ &  $5.9 \times 10^{-11}$   \\ \hline
\end{tabular}
\caption{ 
Benchmark points that satisfy phenomenological constraints.}
\label{tab:2}
\end{table}

 \subsection{Relic density of DM}
 \label{subsec:Relic}
 
\begin{figure}[tb]
\begin{center}
\includegraphics[width=100.0mm]{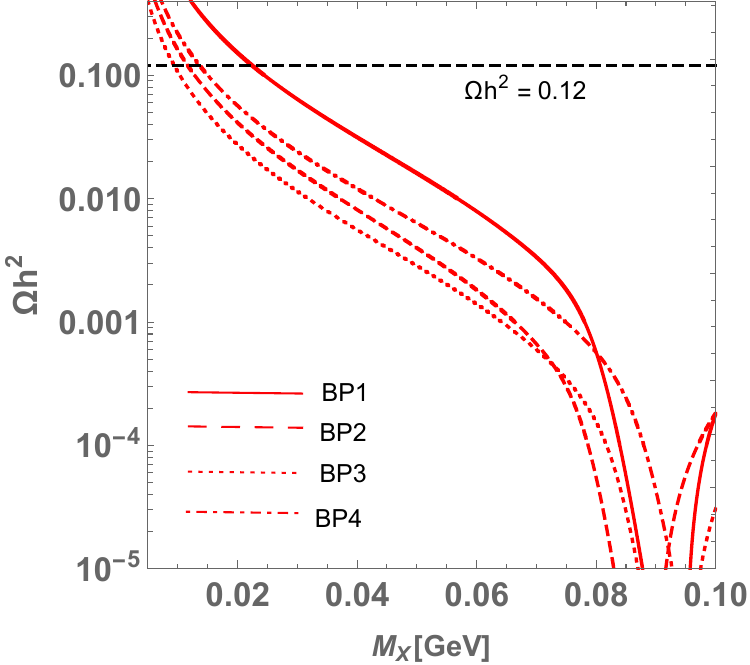} 
\caption{Relic density of DM as a function of mass for each BP.}
  \label{fig:DM1}
\end{center}\end{figure}

 We discuss the relic density of DM applying our benchmark points for new gauge couplings and $Z'_{1,2}$ masses.
 The dominant DM annihilation in our benchmark points is $\bar \chi \chi \to Z'_2 \to \nu \bar \nu$ via $s$-channel, where we consider DM mass less than $0.1$ GeV so as to forbid annihilation into muon and tauon modes kinematically.
We calculate the relic density of DM using {\it MicrOMEGAs 5.2.4}~\cite{Belanger:2014vza} where 
we input relevant interactions and parameters into the code. 
 Here four BPs in Table.~\ref{tab:2} are applied for gauge boson masses and couplings, and we scan DM mass as a free parameter within the range of $[0.005, 0.1]$ GeV.
 In Fig.~\ref{fig:DM1}, we show the relic density of the DM as a function of mass for each BPs. 
The observed relic density $\Omega h^2 = 0.1200 \pm 0.0012$ is shown as black dashed line \cite{Planck:2018vyg}.
We find that observed relic density can be obtained at non-resonant regions due to the sizable value of the $g_H$ coupling.
 
 \subsection{Neutrino signature from DM annihilation}
\begin{figure}[tb]
\begin{center}
\includegraphics[width=120.0mm]{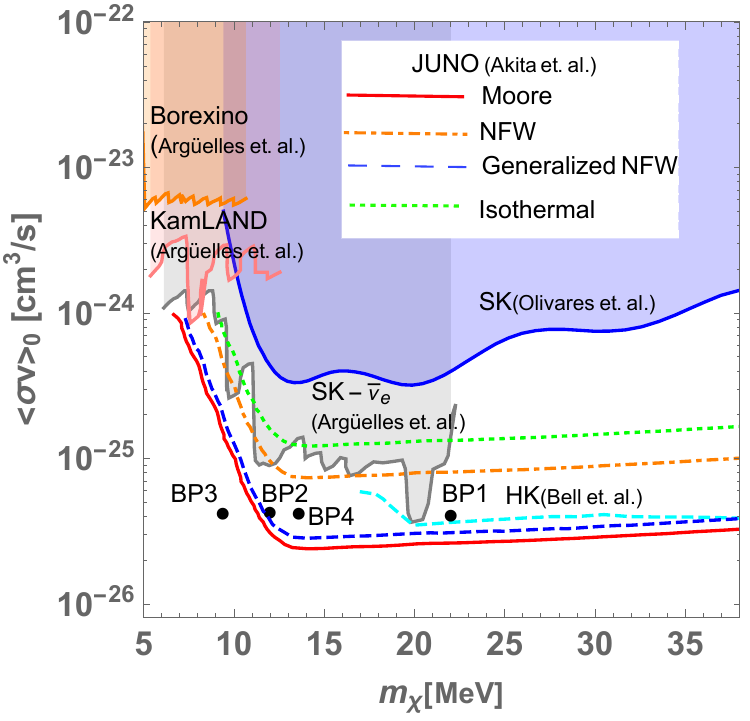}
\caption{ Thermal cross section at the current universe for each BP compared with several experimental constraints and perspective. }
  \label{fig:DM2}
\end{center}\end{figure}

 As discussed in Section \ref{subsec:Relic}, the DM annihilation channel that gives the major contribution at the benchmark points is neutrino pair production. It is the main annihilation channel for the indirect detection of DM as well as relic abundance. Even though the neutrino final state is difficult to detect, the neutrino flux can be monochromatic because the kinetic energy of the neutrino is the same as DM mass in the initial state. The most promising way for detecting such a signal is underground neutrino detectors. Several experiments have already put limits on the current thermally-averaged DM annihilation cross section to neutrinos $\langle \sigma v\rangle_0$.
They give the following constraints~\cite{Palomares-Ruiz:2007trf, Olivares-DelCampo:2017feq, Arguelles:2019ouk}; Borexino $\langle \sigma v\rangle_0 \lesssim 4\times 10^{-24}-10^{-22}$ cm$^3$/s for the DM mass range of [1-10] MeV and Super-Kamiokande (SK) and KamLAND $\langle \sigma v\rangle_0 \lesssim 4\times 10^{-26}-8\times10^{-25}$ cm$^3$/s for the range of [10-10$^3$] MeV.
These constraints are not so stringent, and a wider parameter range is expected to be verified by Hyper-Kamiokande (HK) and JUNO in the future.
Figure \ref{fig:DM2} shows the current and expected constraints to the proposed model from neutrino detection experiments.
The orange, pink, gray and blue-filled regions correspond to constraints by Borexino, KamLAND, SK-$\bar{\nu}_e$ and SK, respectively.
As the next generation experiment, the expected limit by HK for 20 years run time supposing NFW profile \cite{Navarro:1995iw} is shown in the dashed cyan line \cite{Bell:2020rkw}. For detailed profile dependence on HK limits, see Ref.~\cite{Bell:2020rkw}. 
The expected limit by JUNO for 20 years run time is also shown in the red, dot-dashed orange, dashed blue, and dotted light-green lines; each line corresponds to the case of generalized NFW \cite{Benito:2019ngh}, NFW, Moore \cite{Moore:1999gc}, and Isothermal \cite{Bahcall:1980} profiles of dark matter in the Galaxy \cite{Akita:2022lit}.
Even though the dependence on the dark matter profile is considerable, there is a possibility to generally check BP1 with HK.  A similar rough check can be possible for BP1, BP2, and BP4 by JUNO.

\section{Summary and Conclusions}

In this paper, we have discussed a model in which two different extra $U(1)$ gauge symmetries are introduced.
The one is local $U(1)_{L_\mu - L_\tau}$ symmetry and the other one is hidden $U(1)$ symmetry under which the SM fields are neutral.
After spontaneous symmetry breakings of these $U(1)$, we have two $Z'$ bosons that mix each other and they interact with $\mu$ and $\tau$ flavor leptons.
The mass eigenvalues and eigenstates for two $Z'$ bosons are formulated by gauge interactions in mass basis.
We also introduced DM candidate which is Dirac fermion charged under hidden $U(1)$.

We have derived some phenomenological formulas associated with $Z'$ interactions such as muon $g-2$, effective interactions for semi-leptonic $B$ meson decays, $B$ meson decay width for the mode including light $Z'$ and $K^{(*)}$ meson and cross section for neutrino trident process.
Then we have considered a scenario where heavier $Z'$ has a mass of  $\mathcal{O}(10)$ GeV and the lighter $Z'$ has  $\mathcal{O}(10)$ MeV, and  carried out numerical analysis searching for parameter region which is allowed by phenomenological constraints and induce sizable muon $g-2$. 
Some benchmark points have been found that are phenomenologically viable.
For these benchmark points, we have discussed DM physics scanning its mass where DM dominantly annihilates into neutrinos thanks to the gauge mixing and $U(1)_{L_\mu-L\tau}$, and found mass points that explain observed relic density.
Furthermore, we have estimated DM annihilation cross section in the current universe and discussed the possibility of observing neutrino signatures from DM annihilation at neutrino experiments such as HK and JUNO.
It has been found that some benchmark points can be tested at HK and/or JUNO for some DM profiles in the future. 

Before closing our paper, it is worthwhile mentioning the neutrino sector. In fact, our model provides a type ${\bf C^R}$ in Table~1 of ref.~\cite{Asai:2018ocx}, {\it i.e.}, $(m_\nu^{-1})_{22}=(m_\nu^{-1})_{33}\simeq 0$, where $m_\nu$ is neutrino mass matrix that is constructed by canonical seesaw and the neutrino mass texture is determined only via structure of Majorana mass matrix $\nu_R$. Therefore, the Dirac mass matrix and the charged-lepton one are diagonal due to the gauged $U(1)_{L_\mu-L_\tau}$ assignments. The ${\bf C^R}$ leads us to some predictions such as Dirac CP phase, sum of neutrino masses, and the effective mass for neutrinoless double beta decay which has already been discussed in {\it e.g.} ref.~\cite{Asai:2017ryy}.

\section*{Acknowledgments}
The work was supported by JSPS Grant-in-Aid for Scientific Research (C) 21K03562 and (C) 21K03583 (K.~I.~N).
This research was supported by an appointment to the JRG Program at the APCTP through the Science and Technology Promotion Fund and Lottery Fund of the Korean Government. This was also supported by the Korean Local Governments - Gyeongsangbuk-do Province and Pohang City (H.O.). 
H. O. is sincerely grateful for KIAS and all the members.
The work was also supported by the Fundamental Research Funds for the Central Universities (T.~N.).
The work was supported by JSPS KAKENHI Grant Nos.~JP18K03651, JP18H01210, JP22K03622 and MEXT KAKENHI Grant No.~JP18H05543 (T.~S.).

\end{document}